\documentclass[prl,twocolumn,floatfix,showpacs]{revtex4}
\usepackage{graphicx,subfigure,amsmath,colordvi,times}
 \usepackage[usenames]{color}
 \usepackage{multirow}
 
\begin{document}
 \title{High Temperature Superfluidity in Double Bilayer Graphene }
\author{A. Perali$^1$, D. Neilson$^{1,2}$, A.R. Hamilton$^3$}
 \affiliation{$^1$
 Universit\`a di Camerino, 62032 Camerino, Italy\\
 $^2$NEST CNR-INFM, 56126 Pisa, Italy\\
 $^3$
 University of New South Wales, Sydney 2052, Australia}

\begin{abstract}
Exciton bound states in solids between electrons and holes are predicted to form a superfluid at high temperatures.  We show that by employing atomically thin crystals such as a pair of adjacent bilayer graphene sheets, equilibrium superfluidity of electron-hole pairs should be achievable for the first time.  The transition temperatures are well above liquid helium temperatures.  Because the sample parameters needed for the device have already been attained in similar graphene devices, our work suggests a new route towards realizing high-temperature superfluidity in existing quality graphene samples.
\end{abstract}
\maketitle

It is proving a challenging task to observe superfluidity in a semiconductor electron-hole double quantum well system. Despite long-standing theoretical predictions\cite{Lozovik-1975-76,VigMacD96} and significant experimental efforts\cite{Sivan,Croxall,Lilly}, it is only very recently that Bose-Einstein condensation of excitons has been observed in this system.\cite{BECintrap}  The transition temperature was low, $\sim 1$ K, and the condensate was non-equilibrium because of the fast recombination of the photo-excited excitons.
Major obstacles blocking experimental realization of equilibrium
superfluidity in these systems include the following.
(a) In most semiconductors, the electron and hole energy bands are badly mismatched.
In GaAs
not only do the effective masses differ by a factor of five, but the holes have highly non-parabolic energy dispersion and spin-$\frac{3}{2}$ characteristics.\cite{Zulicke}  (b) The large bandgap in GaAs means a bias of $1.5$ eV needs to be applied across the dielectric barrier.  This leads to a huge electric field across the barrier making it extremely hard to fabricate barriers thin enough for the electrons and holes to bind together while
avoiding leakage between the layers. The formation of excitons is exponentially suppressed once the thickness $D_B$ of the barrier separating the two quantum wells exceeds the effective Bohr radius $a_0^\star$\cite{Lozovik-1975-76} ($13$ nm for GaAs).  With the narrowest GaAs quantum wells widths $D_{W}\sim 20$ nm and barrier thicknesses $D_B\agt 10$ nm to avoid leakage, the minimum effective layer separation $D_{eh}\simeq D_{B}+D_{W}$ remains $\gg a_0^\star$.\cite{Croxall}

An alternative system, two graphene monolayers of electrons and holes separated by a dielectric barrier (2xMLG), has recently been proposed to observe this elusive superfluid.\cite{SuperBLG24,MBSM2008,BMSMcomment}
This system has some clear advantages over the GaAs system.  Graphene is a gapless semiconductor with nearly identical conduction and valence bands so the mismatch between the electron and hole Fermi surfaces is almost eliminated.
In addition, the  availability of very thin  dielectric barriers separating the two monolayers makes the region with strong electron-hole pairing effects easily attainable.  A barrier thickness as small as $D_{B}\simeq 1$ nm has already been demonstrated with a hexagonal boron nitride (hBN) dielectric.\cite{Gorbachev}  The barrier can be made so thin for graphene both because there is no need for a large bias between the electron and hole layers, and also because hBN has a much larger bandgap ($\sim 5$ eV) than the barrier between double quantum wells in GaAs ($\sim 0.5$ eV), allowing the barrier thickness to be readily reduced without electrical leakage between the layers.

However there is a new obstacle with graphene associated with the linear single-particle energy dispersion $E_\pm(\mathbf k)=\pm\hbar v_F k$.  This makes it difficult to access the region of strong interactions.  Also, bound excitons do not form because of the massless carriers.\cite{Lozovik2008} The ratio $r_s=\langle V_{Coul}\rangle/E_F$ is a useful measure of the importance of interactions relative to kinetic energy.  $\langle V_{Coul}\rangle = (e^2/ \kappa)\sqrt{\pi n}$  is the average Coulomb interaction energy, where $n$ is the charge carrier density and $\kappa$ is the dielectric constant.  In graphene the Fermi velocity $v_F\sim 10^6$ ms$^{-1}$ is independent of $n$ and the Fermi momentum for the spin-$\frac{1}{2}$ carriers in 2D is $k_F=\sqrt{2\pi n/g_v}$, where  $g_v=2$ is the pseudospin factor.  The Fermi energy is then $E_F=\hbar v_F \sqrt{2\pi n/ g_v}$.  This makes $r_s=e^2 /(\kappa\hbar v_F) \alt 1$  for monolayer graphene, a constant of order unity, while recent theory suggests that an electron-hole superfluid
can
only occur at measurable temperatures for $r_s>2.3$ in the 2xMLG system.\cite{Lozovik2012}  Very recent experiments have shown no evidence of superfluidity in this system despite achieving barrier thicknesses as low as $D_B=1$ nm.\cite{Gorbachev}  This poses an exciting challenge: can  new experimentally realistic structures be designed using atomically thin crystals that  allow the transition to a superfluid state?

\begin{figure}[t]
\includegraphics[width=0.48\textwidth]{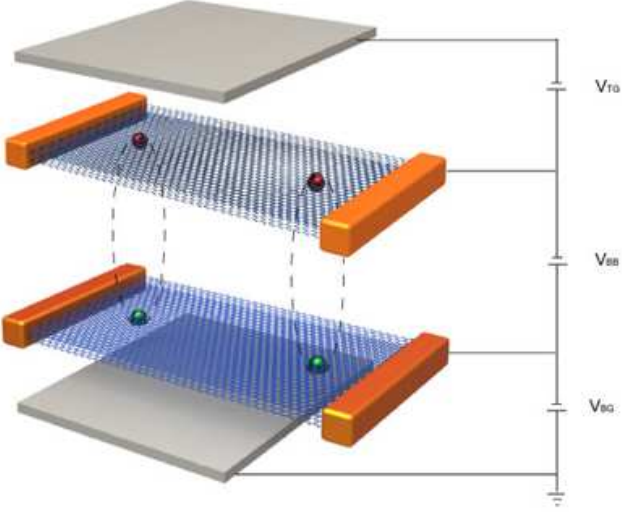}
\caption{Spatially separated electron-hole system (2xBLG) with electrons in one graphene bilayer sheet separated by a hBN dielectric barrier from holes in a second graphene bilayer sheet.   Top and bottom metal gates control the densities.
}
 \label{Fig.1}
 \end{figure}

We concentrate here on bilayer graphene since it has been well characterized and exhibits extremely low levels of disorder, but a number of other such crystals are possible.\cite{Novoselov2005}  Our proposed system
(2xBLG)
consists of a pair of parallel bilayer graphene sheets (Fig.\ \ref{Fig.1}).  The lower bilayer sheet is an electron
bilayer comprising two parallel, $A$-$B$ stacked, closely coupled electron layers of graphene with
layer separation $D_{e}=0.37$ nm.\cite{Nilsson}  There is strong electron hopping between the two layers in the sheet.  The upper bilayer sheet is a hole bilayer consisting of two $A$-$B$
stacked hole layers, but
is otherwise analogous to the lower bilayer sheet with an identical layer separation $D_{h}=D_{e}$.  The two bilayer sheets are separated by a hBN insulating barrier of width $D_{B}$ to prevent tunneling
between the sheets.
There are separate electrical contacts to the two layers and a bias $V_{BB}$ can be applied between them.
The bias $V_{BB}$ and biases $V_{TG}$ and $V_{BG}$ on top and bottom metal gates allow independent control over the carrier density in each layer and can adjust the symmetry of the electric field across the two sheets.  By tuning the three biases, a wide range of electron and hole densities can be achieved.

Bilayer graphene eliminates the problems caused by the linear dispersion of monolayer graphene since over a wide range of electron or hole densities, $1\times10^{11}<n<4\times10^{12}$ cm$^{-2}$, symmetrically biased graphene bilayers behave as a zero gap semiconductor with a quadratic dispersion around the Fermi level\cite{McCannFalkoPRL2006}:
$
E_\pm(\mathbf k)\simeq \pm\hbar^2 k^2/2m^\star.
$
The effective mass is $m^\star \simeq (0.03$ to $0.05) m_e$, depending on the carrier density.\cite{Zou}
At very high densities $n > 4\times10^{12}$ cm$^{-2}$, $E_\pm(\mathbf k)$ crosses over to linear behavior,
but this density region lies in the weakly interacting regime and is not of interest.
At very low densities $n\alt n_{min}=1\times10^{11}$ cm$^{-2}$ a trigonal warping of the bands transforms the two quadratic bands into three sets of Dirac-like linear bands.\cite{CastroNetoRMP2009}
This warping can be reduced if necessary by applying an asymmetric bias across each of the two bilayers which will also open up a gap separating the conduction and valence bands.

With quadratic energy dispersion, the Fermi energy in the bilayer sheet depends linearly on density, $E_F=\pi\hbar^2n/g_v m^\star$, so the parameter $r_s$ now has a density dependence, $r_s=(e^2/\kappa)(g_v m^\star/\hbar^2)(1/\sqrt{\pi n})$.  To experimentally reach large $r_s$ and the strongly interacting regime in the bilayer sheet, it is enough to decrease the carrier density $n$ with suitable gate voltages.
In graphene sheets the lowest density is restricted by onset of electron puddle formation, which occurs at densities below  $10^{11}$ cm$^{-2}$ in high quality  graphene bilayers.\cite{Gorbachev}  This corresponds to $r_s=9$ in the strongly interacting regime.

We know that $D_{eh}\ll a_0^\star$ ensures strong electron-hole pairing.  For a 2xBLG system embedded in a hBN dielectric with $\kappa=3$,\cite{Young} setting $m^\star=m_e^\star\simeq m_h^\star=0.04 m_e$ and using the reduced mass yields an effective Bohr radius $a_0^\star=8$  nm.  This is large compared with the thickness of barriers already fabricated, $D_B\simeq 1$ nm $\simeq D_{eh}$.\cite{Gorbachev}

\begin{table}
\begin{tabular}{||c|c|c|c||c|c|c|c|c||}
\hline
System & $\kappa$   &$a_0^\star$& Ry$^\star$ & $D_{eh}$& $r_{0} $& $D_{eh}/a_0^\star$&$r_s$ \\
\hline
\hline
2xBLG      &  $3$        &$8$ nm    & $ 30$ meV &  $1$ nm & $18$ nm &  $0.1$           & $9$ \\
\hline
2xMLG      &  $3$        & $-{\mathbf{\dagger}}$        &  $-{\mathbf{\dagger}}$      &  $1$ nm & $18$ nm &  $-$              & $0.5$ \\
\hline
GaAs DQW   &  $13$       & $13$ nm    &  $4.5$ meV& $25$ nm & $23$ nm &$2$& $2$  \\
\hline
 \end{tabular}\
\caption{Parameters for 2xBLG, 2xMLG, and electron-hole GaAs double quantum well (GaAs DQW) systems. For $a_0^\star$ and Ry$^\star$ we use the reduced mass. $D_{eh}$ is  minimum effective layer separation. $r_{0}$ and $r_s$ are maximum particle spacing and maximum $r_s$  experimentally attained.
($^{\mathbf{\dagger}}$Localized excitons do not form in 2xMLG.\cite{Lozovik2008})
}
 \label{tablep}
\end{table}
Table I summarizes key physical parameters for three systems in which superfluidity of spatially separated electron-hole pairs has been predicted.  For GaAs DQW, although the $E(\mathbf k)$ is quadratic, the following  experimental restrictions make it difficult to access the region with strong pairing effects. (a) The barriers are wide, $D_{eh}\agt 2a_0^\star$,\cite{Croxall,Lilly} (b) the largest value of $r_s$ attained is $r_s=2$, and (c) the effective Rydberg (Ry$^\star$) binding energy is small.  For 2xMLG, the barriers can be very thin, $D_{eh}\ll a_0^\star$, but the linear $E_\pm(\mathbf k)$ keeps the system in the weakly interacting regime with $r_s< 1$.  What decisively favors the 2xBLG system is that the strong pairing regime is accessible in current samples because:
(a) the extremely thin barriers $D_B\geq 1$ nm and thin bilayer sheets $D_{e}=0.37$ nm, 
(b) the ability to tune $r_s$ to large values in order to access strong pairing,
(c) the almost perfectly matched electron and hole bands resulting from their near equal effective masses, leading to almost perfect particle hole symmetry and nesting between circular Fermi surfaces, and
(d) the larger Ry$^\star$ than for GaAs.

We now calculate the superfluid energy gap and transition temperature to see whether a superfluid state can form in the 2xBLG system with realistic sample parameters and at experimental attainable temperatures.
For simplicity we restrict $D_{B}> D_{e}$ and approximate the system by a single layer of electrons $\ell=e$ interacting with a single layer of holes $\ell=h$.
The quadratic energy bands are $\xi_{\mathbf{k}}^{\gamma\ell}= \gamma\hbar^2{k}^{2}/(2 m_{\ell}^\star) - \mu^{\ell}$, where the band index $\gamma=1$ and $-1$ for the upper and lower bands and $\mu^{\ell}$ is the chemical potential.   We make a standard transformation to positively charged particles for the holes.
The effective Hamiltonian is
\begin{eqnarray}
{\cal{H}}\! =\! \sum_{\ell\mathbf{k}\gamma} \xi_{\mathbf{k}}^{\gamma\ell} c^{\gamma\ell\ \dagger}_{\mathbf{k}} c_{\mathbf{k}}^{\gamma\ell}
 \!+\!\sum_{\substack{\mathbf{q}\mathbf{k}\gamma\\ \mathbf{k}'\gamma'}} V^{eh}_{\mathbf{k}-\mathbf{k}'}
c^{\gamma e\ \dagger}_{\mathbf{k}+\frac{\mathbf{q}}{2}}
c^{\gamma h\ \dagger}_{-\mathbf{k}+\frac{\mathbf{q}}{2}}
c^{\gamma' e}_{\mathbf{k}'+\frac{\mathbf{q}}{2}}
c^{\gamma' h}_{-\mathbf{k}'+\frac{\mathbf{q}}{2}}
\label{Grand-canonical-Hamiltonian}
\end{eqnarray}
$c^{\gamma\ell\ \dagger}_{\mathbf{k}}$ and $c_{\mathbf{k}}^{\gamma\ell}$
are creation and destruction operators for charge carriers in layer $\ell$ and band $\gamma$.  Spin indices are implicit.  $V^{eh}_{\mathbf{k}-\mathbf{k}'}$ is the  static screened electron-hole interaction.

\begin{figure*}
\subfigure[\ ]{
\includegraphics[width=0.48\textwidth] {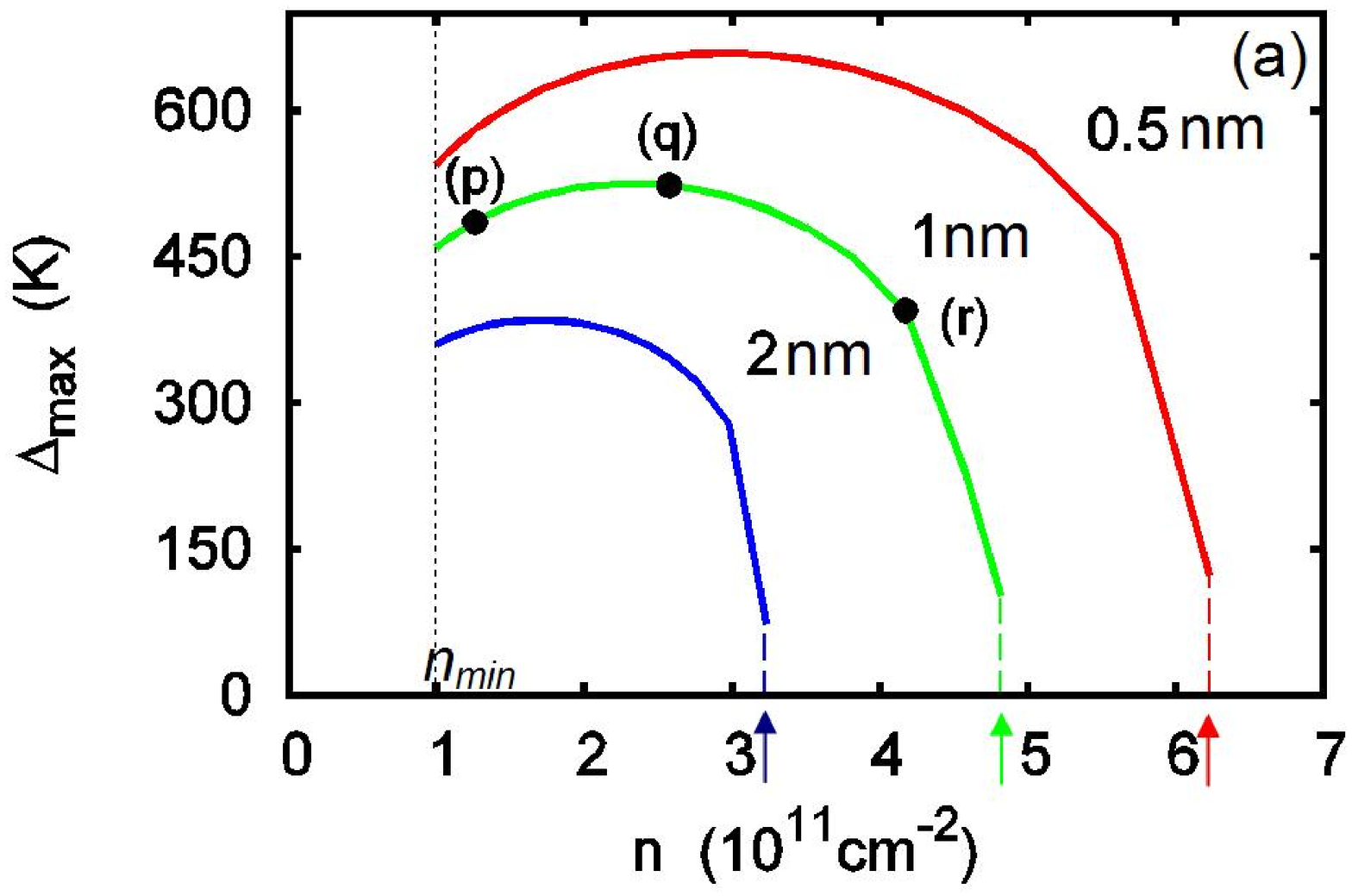}
}
\subfigure[\ ]{
\includegraphics[width=0.48\textwidth] {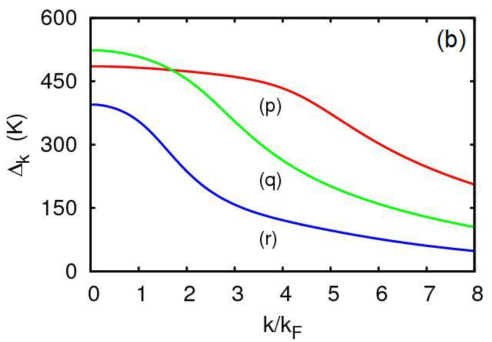}
}
\caption{(a) Maximum of $\Delta_{\bf k}$
at $T=0$ for different barrier thicknesses $D_{eh}$.  Above a critical density (arrows), $\Delta_{\mathrm{max}}$ drops discontinuously to sub-mK energies.  Bands are not quadratic for $n<n_{min}$.  
(b) $\Delta_{\bf k}$ for $D_{eh}=1$ nm at densities $(p)$, $(q)$, and $(r)$ marked in (a)
}
 \label{Fig.2}
 \end{figure*}

A mean-field description is applicable on both the weak-coupling and strong-coupling sides of the BCS-BEC crossover for conventional pairing systems.\cite{Nozieres-Pistolesi-99}  The  mean-field equations at temperature $T$ for  $\mu^{\ell}$ and the momentum-dependent gap function $\Delta_{\mathbf{k}}^{\gamma}$ are,
\begin{eqnarray}
\Delta_{\mathbf{k}}^{\gamma}(T)
 \!=\! - \sum_{\mathbf{k}'\gamma'} \frac{1}{2} V^{eh}_{\mathbf{k}-\mathbf{k}'}
\frac{\Delta_{\mathbf{k}'}^{\gamma'}
}{2 E_{\mathbf{k}'}^{\gamma'}}
[1 - f(E^{\gamma'(+)}_{\mathbf{k}'}) - f(E^{\gamma'(-)}_{\mathbf{k}'})]
\label{Delta-eqn} \\
n_{\left(\substack{e\\h}\right)}(T)\!=\! 2g_v  \sum_{\mathbf{k}\gamma}
[
(u_{\mathbf{k}}^{\gamma})^2 f(E^{\gamma (\pm)}_{\mathbf{k}})
+ (v_{\mathbf{k}}^{\gamma})^2 [1 - f(E^{\gamma (\pm)}_{\mathbf{k}})]
]
\label{n-eh}
\end{eqnarray}
$E^{\gamma (\pm)}_{\mathbf{k}}\!= E_{\mathbf{k}}^{\gamma} \pm \frac{1}{2} (\xi^{{\gamma e}}_{\mathbf{k}} -  \xi^{{\gamma h}}_{\mathbf{k}})$,
$E_{\mathbf{k}}^{\gamma} =\sqrt{\frac{1}{4}(\xi^{{\gamma e}}_{\mathbf{k}}+\xi^{{\gamma h}}_{\mathbf{k}})^2 +\Delta_{\mathbf{k}}^{\gamma\ 2}}$, and $f(E)$ is the Fermi distribution function.
Since we are in the strong coupling regime, we retain only the $s$-wave harmonic in the graphene geometrical form factor.\cite{Lozovik2012}

We calculate the static screened $V^{eh}_{\mathbf{q}}(T)$ within the RPA,
\begin{equation}
\!V^{eh}_{\mathbf{q}}(T)\!=\!\frac{v_q\mathrm{e}^{-qD_{eh}}}{1-2v_q\Pi_0(q,T)+v_q^2\Pi_0^2(q,T)[1-\mathrm{e}^{-2qD_{eh}}]}\!
\label{Vscr}
\end{equation}
$v_q=-2\pi e^2/(\kappa q)$ is the unscreened Coulomb interaction. $\Pi_0(q,T)=\Pi_0^{(n)}(q,T)+ \Pi_0^{(a)}(q,T)$ is the sum of the normal and anomalous polarizabilities of a graphene bilayer sheet, defined in Eqs.\ 6 and 7 of the supplementary material.

There has been considerable discussion on the effect of screening on superfluidity in electron-hole systems. Screening weakens the interactions that drive the superfluid but there is an important subtlety: the appearance of even a small superfluid gap in the excitation energy spectrum would completely suppress small momentum transfer screening.\cite{GortelSwier96,BMSMcomment,Lozovik2012}  In the 2xMLG system, screening suppresses
superfluidity
for all physically accessible parameters,\cite{Lozovik2012} but in the 2xBLG system the interactions are stronger leading to more localized pairs and this can further reduce the effectiveness of screening.  Ultimately only full calculations can reveal whether the reduction is sufficiently great to allow onset of high-$T$ superfluidity, but the experimental observation in Ref.\ \onlinecite{BECintrap} of Bose-Einstein condensation of dilute electrons and holes in GaAs shows that screening does not invariably suppress superfluidity.  Moreover, the fact that the measured transition temperatures in Ref.\ \onlinecite{BECintrap} are consistent with mean-field calculations using unscreened Coulomb interactions, demonstrates that superfluidity can indeed suppress screening.

We solved Eqs.\ \ref{Delta-eqn} to \ref{Vscr} at $T=0$.  Figure \ref{Fig.2}(a) shows $\Delta_{\mathrm{max}}$, the maximum of $\Delta_{\bf k}$ at $T=0$, for different barrier thicknesses  $D_{eh}$.
The electron and hole densities are set equal $n_e=n_h=n$
so only the reduced mass enters the
equations.\cite{PNS}
We take equal effective masses $m^\star=0.04m_e$, noting that a 25\% difference between $m_e^\star$ and $m_h^\star$ \cite{Zou} results in only a 10\% change in the reduced mass and $\Delta_{\mathrm{max}}$.  We restrict our density range $n>n_{min}$ to ensure that $E_F$ lies in the quadratic energy band.  We neglect the small contributions from the negative branches of the bands.

Figure \ref{Fig.2}(a) shows for each $D_{eh}$ there is a critical density $n_c$ above which the gap  $\Delta_{\mathrm{max}}$ is, at most, in the sub-mK energy range.
In realistic disordered systems it is unlikely there would be pairing in this case.
At $n=n_c$ there is a sudden discontinuous jump in $\Delta_{\mathrm{max}}$ to much higher energies.  Reference \onlinecite{Lozovik2012} reported a similar effect for the 2xMLG system but for $r_s>2.35$ only, and this cannot be achieved in the 2xMLG system.  As $D_{eh}$ decreases, the pairing interactions become stronger and the superfluidity persists up to a higher $n_c$.

The nature of the superfluidity can be understood by looking at the $k$ dependence of the gap $\Delta_{\bf k}$.  In the weak-coupled BCS limit of Cooper pairs there would be a peak in $\Delta_{\bf k}$ centered at $k=k_F$, while in the BEC region the peak in $\Delta_{\bf k}$ would be centered at $k=0$ with a long tail falling off as $1/k$, indicating pairs localized in real space.   Figure \ref{Fig.2}(b) shows $\Delta_{\bf k}$ at different densities for $D_{eh}=1$ nm.  The densities $(p)$, $(q)$, and $(r)$  are marked on Fig.\ \ref{Fig.2}(a). At density $(p)$, $\Delta_{\bf k}$ is constant out to $k\sim 4k_F$, indicating an average pair radius comparable to the average particle spacing.  Thus we are close to the BEC region. At densities $(q)$ and $(r)$ the peak is becoming less broad but is still centered on $k=0$ with a long $1/k$ tail.  Further increasing the density should cause a peak at $k_F$ to form but the gap collapses before that happens so the superfluid does not reach the BCS limit before the effectiveness of the screening in Eq.\ \ref{Delta-eqn} discontinuously jumps and superfluidity is suppressed.

The very large gaps shown in Fig.\ \ref{Fig.2}(a), in some cases  $\Delta_{\mathrm{max}}>500$ K, are particularly interesting as, in contrast with high-$T_c$ superconductors, these  values of $\Delta_{\mathrm{max}}$ are maintained across a wide density range.  For reference,  at $n= 5\times 10^{11}$ cm$^{-2}$ the Fermi temperature $T_F=175$ K. 

To determine the superfluid transition temperatures $T_c$, we first recall that for superfluids in 2D the mean-field critical temperature $T_c^{MF}$, the temperature at which the mean-field superfluid gap goes to zero, can overestimate the true $T_c$.  The superfluid transition in 2D systems has a topological Kosterlitz-Thouless (KT) character\cite{KT1973}, and $T_c$ is determined by the KT temperature,
\begin{equation}
T_c=T_{KT} = (\pi/2)\rho_s(T_{KT})\ .
\label{T_KT}
\end{equation}
$\rho_s(T)$ is the superfluid stiffness.
Mean-field theory gives a good estimate of $\rho_s(T)$ for quadratic bands in both the BCS and BEC limits.\cite{MBSM2008}  $\rho_s(0)=E_F/4\pi$ at $T=0$, and $\rho_s(T)$ falls off slowly up to $T\sim \Delta_{\mathrm{max}}$
if $k_F D_{eh}$ is small, generally the case for our parameters.
For $n<n_c$ we can take $\rho_s(T)\simeq\rho_s(0)$, Eq.\ \ref{T_KT} then giving $T_{KT}= E_F/8$.  This is because even at a density as high as $n=6\times 10^{11}$ cm$^{-2}$ $E_F/8 \sim 26$ K, still much less than the $\Delta_{\mathrm{max}}$ shown in Fig.\ \ref{Fig.2}(a).  For $n>n_c$, $\Delta_{\mathrm{max}}$  is extremely small.  Since $\rho_s(T)$ collapses to zero as $T$ becomes larger than $\Delta_{\mathrm{max}}$, $T_c=T_{KT}$ is similarly small for $n>n_c$.

\begin{figure}[t]
\includegraphics[width=0.48\textwidth] {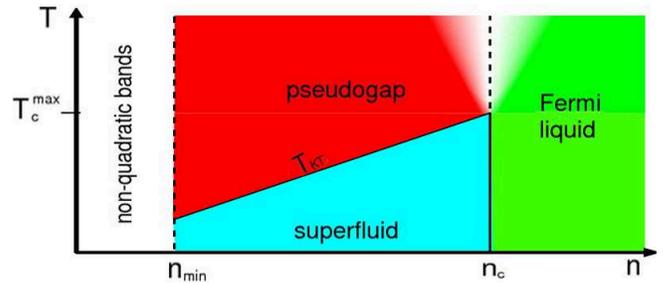}
\caption{Density $n$ -- temperature $T$ phase diagram
(see text).
}
 \label{Fig.3}
\end{figure}

Figure \ref{Fig.3} shows the $T$-$n$ phase diagram.
For densities $n$ greater than a critical density $n_c$, the system is a Fermi liquid at all practicable non-zero temperatures.
For  $n<n_c$ and $T$ below the transition temperature $T_c$, the system is a superfluid.  $T_c= T_{KT}= E_F/8$
grows linearly with density so the maximum transition temperature
$T_c^{max}= E_F(n_c)/8$.
For  $n<n_c$ and $T>T_c$ there will still be strong signatures of the
underlying superfluid state
through the pseudogap.
The pseudogap is a normal state precursor of the superconducting gap due to local dynamic pairing correlations and is produced by non-coherent fluctuations of the pairing field.\cite{Franz}
The deviations from Fermi liquid behavior due to the pseudogap can persist up to room temperature, $T\sim \Delta_{\mathrm{max}}$.\cite{PPSC,GSDJPPS2010}

Table \ref{Table2} gives values of $T_c^{\mathrm{max}}$ and $n_c$ which increase with decreasing $D_{eh}$  because of the stronger electron-hole coupling.  Using hBN barriers between the gates and between the graphene sheets (System (A)) gives $\kappa=3$, and $T_c^{\mathrm{max}}$ is well above liquid helium temperatures.  If the two graphene bilayer sheets are separated by a hBN dielectric barrier and suspended in air between upper and lower gates (System (B)), the effective $\kappa=1.5$ and $T_c^{\mathrm{max}}$ is greatly increased.  Devices similar to System (B) have recently been fabricated with a single bilayer graphene sheet suspended between two gates.\cite{Weitz}  To compare with existing literature\cite{MBSM2008,Lozovik2012} on an idealized and unrealistic configuration of 2xMLG with $\kappa=1$ and $D_{eh}=0$, in 2xBLG the resulting $T_c^{max}=350$ K, the maximum value of $\Delta_{\mathrm{max}}$ is $7500$ K, and $n_c$ corresponds to $r_s=2.8$.
\begin{table}[ht]
\begin{tabular}{||c||c|c||c|c||}
\hline
\ & \multicolumn{2}{c||}{System (A)} & \multicolumn{2}{c||}{System (B)} \\
\hline
$D_{eh}$    &$T_c^{\mathrm{max}}$ (K)  & $n_c$ ($10^{11}$ cm$^{-2}$) &$T_c^{\mathrm{max}}$ (K)& $n_c$ ($10^{11}$ cm$^{-2}$) \\
\hline
$0.5$ nm    & $27$        & $6.2$    &  $72$        &   $17$  \\
\hline
$1$ nm    & $21$          & $4.8$    &  $52$        &   $12$  \\
\hline
$2$ nm    & $14$          & $3.2$    &  $34$        &   $7.9$  \\
\hline
 \end{tabular}
\caption{
Maximum transition temperatures $T_c^{\mathrm{max}}$ and critical densities $n_c$ for barrier thicknesses $D_{eh}$.  System (A) is for the two graphene bilayer sheets embedded in a hBN dielectric.  System (B) is for the two graphene bilayer sheets separated by a hBN barrier and suspended in air between the gates.
}
\label{Table2}
\end{table}

We now discuss the experimental consequences of our phase diagram.
Even at the lowest density $n_{min}=10^{11}$ cm$^{-2}$, the transition temperatures in systems (A) and (B) remain above liquid helium temperatures.
There will be strong experimental signatures of the superfluid below $T_c$.  Although the pairs
are neutral, the ability to make separate electrical contacts to the electron and hole layers allows for spectacular electrical effects.\cite{SuMacDonald2008}  Coulomb drag and inter-layer tunneling measurements will show significant enhancements as $T$ is decreased below $T_c$,\cite{VigMacD96} while counterflow measurements can directly probe the
superflow.\cite{SuMacDonald2008,Eisenstein}
Even up to room temperature there will be strong signatures of the pseudogap in phenomena such as compressibility, specific heat capacity and spin susceptibility. The wide density range over which superfluidity can be observed in a single device  is in marked contrast with high-$T_c$ superconductors, where superconductivity occurs only in a narrow 30\% band of doping centered at optimal doping.

The 2xBLG system is the first multiband system with just one condensate. It is an opposite case to multiband superconductors such as magnesium-diboride where  pairing is only within the bands and
multiple coupled condensates appear, one condensate and one gap for each band.\cite{Peeters}   In contrast, in the 2xBLG system, pairing is only possible between the different hole and electron bands,
leading to a single condensate and a single gap.

The combination of extremely thin barriers, large $r_s$, near-equal effective masses, and strong pairing attraction makes the 2xBLG system ideal for observing high-$T_c$ superfluidity.  In existing quality samples it should be possible to study the phase transition from Fermi-liquid to superfluid as well as the pseudogap physics.
Our approach would also apply more generally to other small gap semiconductors that can be made into atomically thin flakes and used with dielectrics such as hBN.

{\bf
Acknowledgments.} We thank Eva Andrei, Antonio Castro Neto and Pierbiagio Pieri for very useful discussions.

\begin{center}
{\bf Supplementary material.\\ High Temperature Superfluidity in Double Bilayer Graphene}
\end{center}

\begin{center}
A. Perali, D. Neilson, A.R. Hamilton
\end{center}

\section*{Methods}
We have taken the case of equal electron and hole densities $(n_e=n_h=n)$ and equal effective masses $(m^\star=m_e^\star=m_h^\star)$.  For zero temperature, the mean field Eqs.\ (3)-(4) in the main text for the momentum-dependent gap function $\Delta^{\gamma}_{\mathbf{k}}\equiv \Delta^{\gamma}_{\mathbf{k}}(T=0)$ and chemical potentials $\mu^e$ and $\mu^h$ reduce to coupled equations for $\Delta^{\gamma}_{\mathbf{k}}$ and the average chemical potential, $\mu=(\mu_e+\mu_h)/2$,
\begin{eqnarray}
\Delta^{\gamma}_{\mathbf{k}}&=& -\sum_{{\mathbf{k}'}\gamma'} \frac{1}{2}V^{eh}_{\mathbf{k}-\mathbf{k}'} \frac{\Delta^{\gamma'}_{\mathbf{k}'}} {2E^{\gamma'}_{\mathbf{k}'}}
\label{Delta-eqn} \\
n &=& 2g_v\sum_{\mathbf{k}} (v^{\gamma=1}_{\mathbf{k}})^2\ .
\label{n-eh}
\end{eqnarray}
$\xi^{\gamma}_{\mathbf{k}}=\gamma \hbar^2k^2/(2m_r^\star)-\mu$, where $m_r^\star$ is the reduced effective mass of the electron-hole pairs.  The screened zero-$T$ electron-hole Coulomb attraction $V^{eh}_{\mathbf{k}-\mathbf{k}'}$ depends on the momentum exchanged in the scattering process, thus making the gap momentum-dependent.  The extra factor of $\frac{1}{2}$ in Eq.\ (6) comes from the graphene geometrical form factor $F^{\gamma\gamma'}_{\mathbf{k}\mathbf{k}'}$.  In the phase space region of interest to us, coupling is strong and the $s$-wave harmonic contribution to $F^{\gamma\gamma'}_{\mathbf{k}\mathbf{k}'}$ dominates, leading to the factor $\frac{1}{2}$.

The numerical sums are transformed into continuous integrals and then numerically evaluated using the Gauss-Legendre method.  We numerically solved the $T=0$ non-linear system of integral equations, Eqs.\ (6) and (7), iteratively  using a multivariable Newton method.  The convergence ratio requirement for the variables was set at $10^{-6}$.  Retaining only the s-wave harmonic of the form factor, results in a gap function $\Delta_{\mathbf{k}}$ that depends only on the modulus of the momentum ${\mathbf{k}}$.

Concerning the ultraviolet behavior of the mean field equations, the  integral over the momentum is well-defined, thanks to the exponential decay in momentum space of the bare Coulomb interaction $v_{\mathbf{q}}\mathrm{e}^{-qD_{eh}}$ between electrons and holes in the bilayers separated by a dielectric barrier of thickness $D_{eh}$.  In our calculations we fixed the upper momentum cutoff for the integrations as $k_{c}=5\times\max(1/D_{eh},k_F)$.   Numerical stability was checked by increasing $k_c$ and increasing number of integration points.  We confirmed that the results obtained by the Newton method for $\mu>0$ agreed with a solution of the mean field equations independently obtained by a direct recursive method.\\

\section*{Suppression of screening: polarization bubble}

In the superfluid state at low temperature, the static limit of the polarization bubble in the diagrammatic (RPA-like) resummation is  suppressed at small momenta $q$. Reference 13 in the main text discusses this effect in the limiting case of no dielectric barrier, $D_{eh}= 0$, for energy bands with linear dispersion.  The polarization bubble  is responsible for renormalizing the bare Coulomb interaction. A small-$q$ suppression of the screening allows strong unscreened electron-hole pairing peaked at small-$q$ to occur, and this strong pairing can lead to a large gap in the excitation spectrum.

We performed  numerical calculations for the polarization bubbles for the case of quadratic energy bands. In the superfluid state the polarization bubble is given by the sum of the normal and anomalous bubbles (Eq.\ 4 in the main text and supplement Fig.\ 4).

\begin{figure}
\includegraphics[angle=0,width=0.45\textwidth] {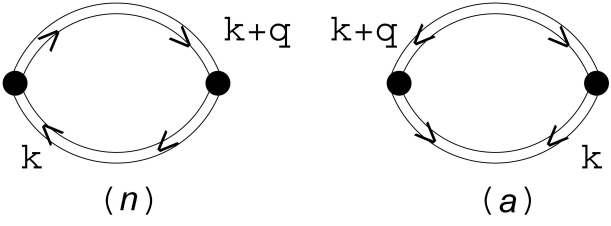}
\caption{$\Pi_0^{(n)}(q)$ and $\Pi_0^{(a)}(q)$ with two normal and two anomalous Green functions, respectively.}
 \label{polbubbles}
 \end{figure}

In RPA the polarization bubbles are constructed with the normal and
anomalous Green's functions of BCS theory.
Assuming the Green's functions are diagonal in the band indices, the normal and anomalous Green's functions are given by,
\begin{eqnarray}
{\cal{G}}^{\gamma\gamma}(\mathbf{k},\mathrm{i}\epsilon_n)
&=&\frac{(u_{k}^{\gamma})^2}{\mathrm{i}\epsilon_n-E_{k}^{\gamma}}
+\frac{(v_{k}^{\gamma})^2}{\mathrm{i}\epsilon_n+E_{k}^{\gamma}}\\
{\cal{F}}^{\gamma\gamma}(\mathbf{k},\mathrm{i}\epsilon_n)
&=&\frac{u_{k}^{\gamma}v_{k}^{\gamma}}{\mathrm{i}\epsilon_n-E_{k}^{\gamma}}
-\frac{u_{k}^{\gamma}v_{k}^{\gamma}}{\mathrm{i}\epsilon_n+E_{k}^{\gamma}}
\ ,
\label{GandF}
\end{eqnarray}
respectively, where $u_{k}^{\gamma}$ and $v_{k}^{\gamma}$ are the Bogoliubov factors
\begin{eqnarray}
(u_{k}^{\gamma})^2&=&\frac{1}{2}\left(1+\frac{\xi_k^\gamma}{E_k^\gamma}\right)\nonumber\\
(v_{k}^{\gamma})^2&=&\frac{1}{2}\left(1-\frac{\xi_k^\gamma}{E_k^\gamma}\right)\ .
\label{uandv}
\end{eqnarray}
The summation
of the fermionic Matsubara frequencies $\epsilon_n$ of the particle-hole loop can then be
done analytically, and taking the static limit the expressions for the
bubbles are given by,
\begin{eqnarray}
\Pi_0^{(n)}(q,T)\!\!&=&\!\! -2g_v\sum_{\mathbf{k}\gamma\gamma'}
\frac{1}{2}\frac{(u_{k}^{\gamma})^2(v_{{\mathbf{k}}-{\mathbf q}}^{\gamma'})^2
+(v_{k}^{\gamma})^2 (v_{{\mathbf{k}}-{\mathbf q}}^{\gamma'})^2} {E_{\mathbf{k}}^{\gamma}+E_{{\mathbf{k}}-{\mathbf q}}^{\gamma'}}\nonumber\\
\label{pi-n} \\
\Pi_0^{(a)}(q,T)\!\!&=&\!\! 2g_v\sum_{\mathbf{k}\gamma\gamma'}
\frac{1}{2}\frac{2u_{\mathbf{k}}^{\gamma} v_{\mathbf{k}}^{\gamma}
u_{{\mathbf{k}}-{\mathbf q}}^{\gamma'} v_{{\mathbf{k}}-{\mathbf q}}^{\gamma'}}{E_{\mathbf{k}}^{\gamma} +E_{{\mathbf{k}}-{\mathbf q}}^{\gamma'}}\ .
\label{pi-a}
\end{eqnarray}
When the chemical potential $\mu$ is positive, the contributions
of the lower band $(\gamma$ or $\gamma'=-1)$ to the total
polarization bubbles are small  because the corresponding $k$-states are filled and far in energy from $\mu$.  Hence they contribute little to the particle-hole processes. We considered only the $\gamma=\gamma'=1$ contribution and, as in the gap equation, only the s-wave harmonic of the
geometrical form factor of graphene $F^{\gamma\gamma'}_{\mathbf{k},\mathbf{k}-\mathbf{q}}$.

Physically, the effect of screening is  suppressed because of the opening of the energy gap $\Delta$ at the Fermi surface.  This exponentially suppresses particle-hole processes with energies less than the gap energy $\Delta$ and it is precisely these low-energy processes that are needed to screen the long-range Coulomb interaction. Once a gap appears in the excitation spectrum, screening exactly vanishes in the $q=0$ limit. From a diagrammatic point of view, in addition to a suppression of the bubble with diagonal Green functions at low momentum, there is an additional canceling contribution from the anomalous bubble with off-diagonal Green functions.  At $q=0$ the cancelation is exact for any non-zero $\Delta$.

\end{document}